# Convective Amplification of Electromagnetic Ion Cyclotron Waves From Ring-distribution Protons in the Inner Magnetosphere

Manish Mithaiwala, Chris Crabtree, Gurudas Ganguli, Leonid Rudakov[1] and Kunihiro Keika[2]

Plasma Physics Division, Naval Research Laboratory, Washington, DC 20375-5346

[1] Icarus Research Inc., P.O. Box 30870, Bethesda, MD 20824-0780, United States

[2] New Jersey Institute of Technology, Newark, NJ 07102, United States.

The growth of electromagnetic ion cyclotron waves (EMIC) due to a ring distribution of Hydrogen ions is examined. Though these distributions are more commonly implicated in the generation of equatorial noise, their potential for exciting EMIC waves is considered here. It is shown that since the ring distribution is non-monotonic in perpendicular velocity, the amplification achieved by this instability is greater than bi-Maxwellian distributions for typical anisotropies, because the waves can maintain resonance over a much longer part of its trajectory. For ring speeds $(V_r)$ close to the Alfven speed $(V_A)$, the growth rate is maximum at parallel propagation but is much larger at oblique angles compared with a bi-Maxwellian, and can have a second peak approximately at $(k_\perp c / \omega_{pH}) V_r / V_A \sim 2.3$ for ring speeds about the parallel thermal speed. Strong wave gain is achieved for moderate ring speeds $(V_r \sim V_A)$. The analysis suggests that EMIC wave activity should be closely associated with equatorial noise.

## 1. Introduction

Electromagnetic ion cyclotron waves (EMIC) waves are believed to be responsible for ring current loss by wave-particle interaction [Fok et al., 1996; Spasojević et al., 2003; Xiao et al., 2011; see also references from Keika et al., 2013]. They may also have a role in the acceleration and loss of relativistic electrons [Summers and Thorne, 2003; Jordanova et al., 2008; see also references from Keika et al., 2013].

Theoretical studies of the convective amplification of electromagnetic ion cyclotron (EMIC) waves have demonstrated that the path integrated wave gain increases with L shell [Horne and Thorne, 1993, 1994]. The gain becomes sufficiently large at L>7, thus accounting for outer magnetosphere observation of EMIC waves. Analysis of THEMIS data [Min et. al., 2012; Usanova et al., 2012] has verified that convective amplification of EMIC waves occurs mostly in the outer magnetosphere beyond L=7. Their findings were in general agreement with results reported from the AMPTE/CCE spacecraft [Anderson et al., 1990, 1992a,b; Keika et al., 2013]. EMIC waves are also observed in the inner magnetosphere (L<5) [e.g., Horne and Thorne, 1993; Erlandson and Ukhorskiy, 2001], particularly under geomagnetically disturbed conditions [Keika et al., 2013]. However, the origin of inner magnetospheric EMIC waves remains unknown. The path integrated wave gain of EMIC waves due to a hot anisotropic ring current modeled with a bi-Maxwellian distribution was shown to be insufficient to reach observable levels [Horne and Thorne, 1993]. The problem is that wave growth only occurs during the initial part of the ray path when the wave normal angle is small.

Several physical mechanisms have been proposed to account for the observations of inner magnetospheric EMIC waves. 1) Ray tracing shows that the density gradient of the plasmasphere boundary may act to refract the wave normal angle of the ray path such that they grow to sufficient amplitudes [Horne and Thorne, 1993]. Indeed EMIC waves have been observed near the plasmasphere boundary [Fraser et al., 1996]. The same would also be true near the magnetopause where the field line compression due to the solar wind may refract the wave such that wave gain could be enhanced [Anderson and Hamilton, 1993]. 2) Inner magnetosphere waves may also originate at slightly higher L values, perhaps outside and adjacent to the plasmapause where enhanced gain is possible, and refract into the inner magnetosphere. It is thought that this process could only explain the lower frequency portion of the wave spectrum [Horne and Thorne, 1993]. 3) Here it is proposed that the wave growth is a result of a perpendicular velocity ring distribution rather than an anisotropic bi-Maxwellian distribution, which were used in the studies described in 1) and 2) above. This model has several advantages: It does not need to be confined to the plasmapause boundary, the growth rate is larger for the same density of the anisotropic distribution, and the wave can maintain resonance over a longer period of its trajectory leading to enhanced gain.

Ring distributions can be produced by substorm injection events over a broad spatial region between noon and premidnight during non-storm times; during storms they may appear as a consequence of quasi-steady injection into the ring current driven by a global convection electric field [Chen et al., 2011]. Proton ring distributions were observed with in-situ measurements in the inner magnetosphere under geomagnetically quiet conditions [e.g., Smith et al., 1973; Williams, 1981; Kistler et al., 1989] and at substorm-associated injections [e.g., Chen et al., 2011]. The distributions are mostly due to the fact that drift paths of energetic protons (>10 keV) injected from the plasma sheet into the inner magnetosphere differ with proton kinetic energies and equatorial pitch angles. Ring distributions have also been successfully reproduced by global ring current models [e.g., Fok et al., 1993]. They are most commonly believed responsible for the generation of magnetosonic waves, often referred to as equatorial noise [Meredith et al., 2008].

In the Section 2 we consider the instability of EMIC waves due to a hydrogen ring velocity distribution. Then in Section 3 the gain is determined by computing the growth rate along the ray path. Section 5 gives our conclusions about the importance of a ring distribution to EMIC wave generation. We also discuss the applicability of these results to current satellite missions and future modeling efforts.

**2. Linear Instability of EMIC waves from a Hydrogen Velocity Ring Distribution**

The instability for EMIC waves due to a velocity ring distribution of hydrogen ions in a cold magnetospheric plasma applicable to the inner magnetosphere is considered here. The cold plasma ions consist of hydrogen, helium, and oxygen. This magnetospheric equilibrium is perturbed by a perpendicular velocity-ring distribution, which can be modeled as a perpendicular drifting Maxwellian distribution

$$f_{ring}(v_\perp) = \frac{C}{2\pi^{3/2} v_{tr\perp} V_r} \exp\left(-\frac{(v_\perp - V_r)^2}{v_{tr\perp}^2}\right), \tag{1}$$

normalized such that $\int 2\pi v_\perp dv_\perp f_{ring}(v_\perp) = 1$, and $C=1$ when $V_r \gg v_{tr\perp}$. The distribution is peaked at the ring speed $v_\perp = V_r$ with perpendicular thermal speed $v_{tr\perp}$, where parallel (∥) and perpendicular (⊥) are with respect to the direction of the background magnetic field $B_0$. This establishes an important distinction between typical loss cone distributions, modeled for instance by a subtracted Maxwellian (Denton et al., 1992), and the ring-distribution in that there are essentially no particles for a broad range of small $v_\perp$ and not just near $v_\perp = 0$ as in a subtracted Maxwellian distribution. From observations, the ring velocity of interest here is about the same as the Alfven velocity [Boardsen et al., 1992; Chen et al., 2011]. The total distribution is $f_{ring}(v_\parallel, v_\perp) = f_{ring}(v_\parallel) f_{ring}(v_\perp)$. The parallel distribution function, $f_{ring}(v_\parallel)$, is considered to be a Maxwellian with parallel thermal speed $v_{tr\parallel}$, though even a 3-D shell distribution can lead to instability [Korablev and Rudakov, 1968].

The dispersion relation for low frequency electromagnetic cyclotron waves is determined from

$$D = \left(\epsilon_{xx} - n_z^2\right)\left(\epsilon_{yy} - n^2\right) - \epsilon_{xy}^2, \tag{2}$$

such that the electric field is left-circular polarized at $k_\perp = 0$, and where it was assumed that $k_\perp^2 c^2 / \omega_{pH}^2 \ll m_H / m_e$. As will be seen in the next section, this constraint on $k_\perp$ is not too restrictive. Of course at oblique angles, this dispersion must be generalized to include the $\varepsilon_{zz}$ component of the dielectric tensor, which physically means that the parallel component of the electric field is no longer ignorable. However a ring distribution instability for low frequency inertial shear Alfven waves when $k_\perp^2 c^2 / \omega_{pH}^2 \gg m_H / m_e$ is indeed possible (Ganguli et al., 2007).

The ion components of the dielectric tensor $\varepsilon$ can be determined by standard methods (Stix, 1992) and are given by

$$\begin{aligned}
\varepsilon_{xx} &\approx \sum_{s=H,He,O} \frac{\omega_{ps}^2}{\Omega_s^2 - \omega^2} \frac{2\Gamma_0(\lambda_s)}{\lambda_s} + \frac{\omega_{pr}^2}{\omega^2} \frac{n^2 \Omega_r^2}{k_\perp^2} \left[ (-1 - \zeta_n Z + \zeta_0 Z) \frac{\partial J_n^2}{V_r \partial v_\perp} - \frac{V_r^2}{v_{tr\parallel}^2} \frac{J_n^2}{V_r^2} Z' \right] \\
\varepsilon_{xy} &= -\varepsilon_{yx} \approx \sum_{s=H,He,O} i \frac{\Omega_s}{\omega} \frac{\omega_{ps}^2}{\Omega_s^2 - \omega^2} \frac{2\Gamma_0(\lambda_s)}{\lambda_s} + i \frac{\omega_{pr}^2}{\omega^2} \frac{n^2 \Omega_r^2}{k_\perp^2} \left[ (-1 - \zeta_n Z + \zeta_0 Z) \frac{\partial}{V_r \partial v_\perp} \left( v_\perp \frac{\partial J_n^2}{2 \partial v_\perp} \right) - \frac{V_r^2}{v_{tr\parallel}^2} \frac{J_n \partial J_n}{V_r \partial v_\perp} Z' \right] , \\
\varepsilon_{yy} &\approx \sum_{s=H,He,O} \frac{\omega_{ps}^2}{\Omega_s^2 - \omega^2} \frac{2\Gamma_0(\lambda_s)}{\lambda_s} + \frac{\omega_{pr}^2}{\omega^2} \frac{n^2 \Omega_r^2}{k_\perp^2} \left[ (-1 - \zeta_n Z + \zeta_0 Z) \frac{\partial}{V_r \partial v_\perp} \left( v_\perp^2 \left( \frac{\partial J_n}{\partial v_\perp} \right)^2 \right) - \frac{V_r^2}{v_{tr\parallel}^2} \left( \frac{\partial J_n}{\partial v_\perp} \right)^2 Z' \right]
\end{aligned} \quad (3)$$

where the integral over $v_\perp$ was done by parts and a delta function approximation was used for the ring distribution $f_{ring}(v_\perp) = \frac{1}{2\pi v_\perp} \delta(v_\perp - V_r)$, which is valid when $V_r \gg v_{tr\perp}$. The derivatives of Bessel functions in (3) are to be evaluated at $v_\perp = V_r$. The plasma and cyclotron frequencies for each species is defined as $\omega_{ps}^2 = 4\pi n_s e^2 / m_s$, and $\Omega_s = \pm eB_0 / m_s c$ for ions(+) and electrons(-), respectively, with density $n_s$, mass $m_s$, and electric charge e. Throughout this article it is assumed that the ring distribution is a small perturbation to the background and quasineutrality is maintained between ion and electron species. It was also assumed in these equations that $V_r^2 \gg v_{tr\perp}^2$, and $n^2 = k^2 c^2 / \omega^2 \gg 1$. The Alfven speed is $V_A = c\Omega_H / \omega_{pH}$. In the limit that $V_r \to 0$, the ring distribution (1) becomes Maxwellian, and the components of the dielectric tensor reduce to that of a bi-Maxwellian plasma with temperature $T_{r\perp,\parallel} = m_s v_{tr\perp,\parallel}^2$. In the limit that $k_\parallel \to 0$, the components of the dielectric tensor reduce to that given by Sharma and Patel [1986]. The plasma dispersion function Z has argument $\zeta_n = (\omega - n\Omega_r)/k_\parallel v_{tr\parallel}$, the Bessel function $J_n$ has argument $k_\perp v_\perp/\Omega_r$, $\Gamma_n(\lambda_s) = \exp(-\lambda_s) I_n(\lambda_s)/\lambda_s$, $I_n$ is the modified Bessel function, and $\lambda_s = k_\perp^2 v_{ts}^2 / 2\Omega_s^2$.

The growth rate can be analytically determined from the dispersion relation (3). Since the velocity ring distribution is only a small perturbation to the cold plasma components, the growth rate is approximately $\gamma = -D_I \big/ \dfrac{\partial D_R}{\partial \omega}$, and is proportional to $\gamma/\omega \propto n_r/n_e$, where $D_R$ and $D_I$ denotes the real and imaginary parts of the dispersion relation respectively. The real part of the dispersion relation gives the cold plasma dispersion relation for EMIC waves (e.g. Mithaiwala et al., 2007).

At parallel propagation the growth rate due to a ring distribution can be obtained from (3) for small $k_\perp$, i.e. $k_\parallel^2 \approx k^2$, in the limit that $k_\parallel v_{tr\parallel} \gg \gamma$ as

$$\frac{\gamma}{\omega} \sim \sqrt{\pi}\, \frac{n_r/n_e}{2|k_z|^2 c^2/\omega_{pH}^2} \left[ \frac{V_r^2}{v_{tr\parallel}^2} \frac{\Omega_r - \omega}{\Omega_r} - 1 \right]. \tag{4}$$

Clearly there is an instability threshold dependent on $V_r$ determined by setting $\gamma = 0$. For nearly perpendicular propagation, $k_\perp^2 \gg k_\parallel^2$, and for $\omega - \Omega_R > |k_\parallel| v_{tr}$ the growth rate is

$$\frac{\gamma}{\omega} \sim \sqrt{\pi}\, \frac{n_r/n_e}{2|k_z|^2 c^2/\omega_{pH}^2} \left\{ \frac{\Omega_r - \omega}{\Omega_r} \frac{V_r^2}{v_{tr\parallel}^2} \frac{J_1^2(\sigma)}{\sigma^2} - \frac{\partial J_1^2(\sigma)}{2\sigma\, \partial \sigma} \right\}, \tag{5}$$

where $\sigma \equiv k_\perp V_r / \Omega_r$ is the argument of the Bessel function evaluated for $v_\perp = V_r$. There is no stability threshold since $\partial J_1^2(\sigma)/\partial \sigma^2$ can become negative, e.g. when $\sigma > 1.9$. Thus, it is possible to generate oblique EMIC waves even when $V_r$ is small. For an anisotropic bi-Maxwellian distribution instability the growth rate at parallel propagation is approximately [Kennel and Petschek, 1966; Cornwall and Schultz, 1970]

$$\frac{\gamma}{\omega} \sim \sqrt{\pi} \frac{n_r/n_e}{2|k_z|^2 c^2/\omega_{pH}^2} \left[ \frac{T_{r\perp}}{T_{r\|}} \frac{\Omega_r - \omega}{\Omega_r} - 1 \right]. \tag{6}$$

The growth rates (4) and (6) are identical if the substitution $T_{r\perp} \rightleftarrows V_r^2, T_{r\|} \rightleftarrows v_{tr\|}^2$ is made.

For magnetospheric application Figure 1 shows the numerically computed growth rate from (2) for several values of the ring speed, and compares it to the growth rate due to a bi-Maxwellian distribution with a temperature anisotropy $A \equiv (T_{r\perp}/T_{r\|}) - 1 > 0$. At oblique angles, because of the contribution of $J_1 \partial J_1(\sigma)/\partial \sigma$ in the $\varepsilon_{xx}$, $\varepsilon_{xy}$ components of the dielectric tensor, the growth rate exhibits a second maximum approximately at $(k_\perp c/\omega_{pH}) V_r/V_A \sim 2.3$ when $\partial J_1^2(\sigma)/\partial \sigma^2$ becomes nearly maximally negative. For magnetospheric applications, $V_r^2/v_{tr\|}^2 \gg 1$, and only terms with coefficient $V_r^2/v_{tr\|}^2$ in the dielectric tensor (3) contribute to the growth rate. Overall the results show that the growth rate from the ring distribution is larger than a bi-Maxwellian if $V_r^2/v_{tr\|}^2 > T_{r\perp}/T_{r\|}$.

## 3. Ray Tracing of the Unstable EMIC waves and the Plasma Model

Though the growth rate maximizes for parallel propagating waves, it should not be concluded that this property alone determines the wave spectrum. The wave spectrum resulting from the ring instability is determined from the path integrated wave gain $G = exp\left(2\int \gamma dt\right)$. Analysis of whistler growth using THEMIS data shows that a gain of $G > 6$ is sufficient for strong wave growth from a background of noise and even less for moderate wave amplitudes [Li et al., 2009]. The wave gain is computed by solving for the growth rate from the dispersion relation (2) at every point over the ray path. A three-dimensional ray-tracing scheme is used to compute the ray paths of the unstable waves [Crabtree et al., 2012a; 2012b]. The background magnetic field is assumed to be dipolar and the background plasma is cold, which includes hydrogen, helium, and oxygen. The plasma density model is the same as in Bortnik et al. [2011], and is based on the diffusive equilibrium of Angerami and Thomas [1964] and similar to that given by Inan and Bell [1977], which has been used for other studies of convective gain of ion cyclotron waves [Horne and Thorne, 1993]. It includes an ionosphere, plasmapause located at L=3, and a number of field aligned density structures [Bortnik et al., 2011]. Figure 2 shows a typical raypath initialized in the equatorial plane at L=3 with frequency $\omega \approx 0.68\Omega_H$ and $k_\perp = 0$. The ray path is traced over the contours of constant density used in the model calculation.

Figure 3 shows the ray trajectory of an initially parallel propagating EMIC wave. The EMIC waves are initialized at $k_\perp = 0$ (corresponding to the maximum growth rate from figure 1). The waves refract as they travel, mostly parallel to the field line, through the inhomogeneous magnetosphere, and $k_\perp$ increases (Figure 3c). Due to the small amount of magnetospheric helium, when the wave frequency matches the hydrogen-helium bi-ion hybrid resonance (or Buchsbaum resonance),

$$\omega_B^2 = \frac{\omega_{pH}^2 \Omega_{He}^2 + \omega_{pHe}^2 \Omega_H^2}{\omega_{pH}^2 + \omega_{pHe}^2}, \tag{7}$$

the waves reflect, i.e. $k_\parallel$ passes through zero and the wave turns around, only to reflect again at the conjugate point in the other hemisphere (see also Mithaiwala et al., 2007 and references therein). Though most of the gain is achieved during its initial transit away from the equator. As stated previously, when $k_\perp$ becomes sufficiently large the EMIC dispersion (3) must be generalized to include $\varepsilon_{zz} \approx -\frac{\omega_{pe}^2}{\omega^2}\zeta_e^2 Z'(\zeta_e) - n_\perp^2$, which accounts for the Landau damping of the waves toward oblique angles. Only the n=1 cyclotron resonance for instability is considered, and the n=0 contribution gives damping. The background electron temperature is chosen to be $T_e = 2eV$ and $\zeta_e = \omega/k_\parallel v_{te}$. Since the ray spends very little time at very oblique angles during the first few equatorial crossings, this term contributes very little overall (see figure 3c). Though perhaps, as is sometimes the case for whistler waves, collisional damping may be more significant than Landau damping [Crabtree et al., 2012b].

Since it is well established that the proton ring distribution is unstable to oblique $(k \sim k_\perp)$ magnetosonic waves with frequency $\omega$: $k_\perp c \Omega_H / \omega_{pH} \approx (n+1/2)\Omega_H$ satisfying the cyclotron resonance condition $v_\parallel = \omega/k_\parallel (1 - n\Omega_H/\omega)$ for 1<n<40, the resonant parallel thermal speed satisfies $v_{tr\parallel}/V_A = (k/k_\parallel)/(2n+1)$. The parallel thermal speed is chosen so that $V_A^2/v_{tr\parallel}^2 = 10$ and $V_r/V_A = \{1.0, 1.5\}$, with a density of $n_r/n_e = \{0.1, 0.15\}$ respectively. For the anisotropy ratio, A=2 is chosen for comparison with previous studies, so that $v_{tr\perp}^2/v_{tr\parallel}^2 = 3$ [Horne and Thorne, 1993]. Figure 3b shows the gain from the ring and bi-Maxwellian distributions. The gain from the velocity-ring distribution with $V_r/V_A = 1.5$ is large enough to achieve strong growth. For smaller ring speeds, the gain is less, but still much greater than that of an anisotropic bi-Maxwellian, which has virtually no gain. More importantly, the maximum gain from the bi-Maxwellian is achieved by the time $k_\perp c/\omega_{pH} \approx 1.5$, while the EMIC resulting from the ring distribution continues until $k_\perp c/\omega_{pH} \approx 2$.

## 4. Discussion/Conclusion

This article has only considered instability due to a ring distribution of protons and instability above the helium cyclotron frequency. The results considered here can readily be extended to a ring distribution of helium ions (Fu et al., 2001; Yamauchi et al., 2012), or to unstable frequencies below the helium cyclotron frequency (Kozyra et al, 1984; Horne and Thorne, 1994). The helium band events are particularly important for the inner magnetosphere (Keika et al., 2013). To realistically consider EMIC waves due to a ring distribution of helium or frequencies below the cyclotron frequency, oxygen must be included (Omidi et al., 2013). Partly because even a small amount of oxygen modifies the dispersion relation, and also because it enables the reflection from the bi-ion resonance of oxygen and helium. Additionally collisional damping should be included in the modeling [Mithaiwala, 2007; Crabtree 2012b].

EMIC wave growth in the inner magnetosphere region is more likely a result of a velocity-ring distribution of ions than from an anisotropic bi-Maxwellian. This has the advantage that the velocity-ring distribution need not be confined near the plasmapause to achieve any gain. Such EMIC waves of ring distribution origin can be amplified in low-density regions and without density gradient. Any refraction that keeps $k_\perp$ small would result in larger amplitude EMIC waves. Indeed ring distributions are observed between *3<L<7* and with a wide range of energies [Smith et al., 1973; Williams, 1981; Kistler et al., 1989; Meredith et al., 2008; Chen et al., 2011; Yamauchi et al., 2012]. The possible relationship between ring distributions and EMIC wave excitement needs to be examined with simultaneous wave and particle measurements with high time and energy resolutions. The analysis presented in this paper highlights a unique opportunity for missions such as the Van Allen Probes.

**Acknowledgements**:

Supported by Naval Research Laboratory base program. The authors would like to acknowledge the RBSPICE team of the Van Allen Probe mission for valuable discussions.

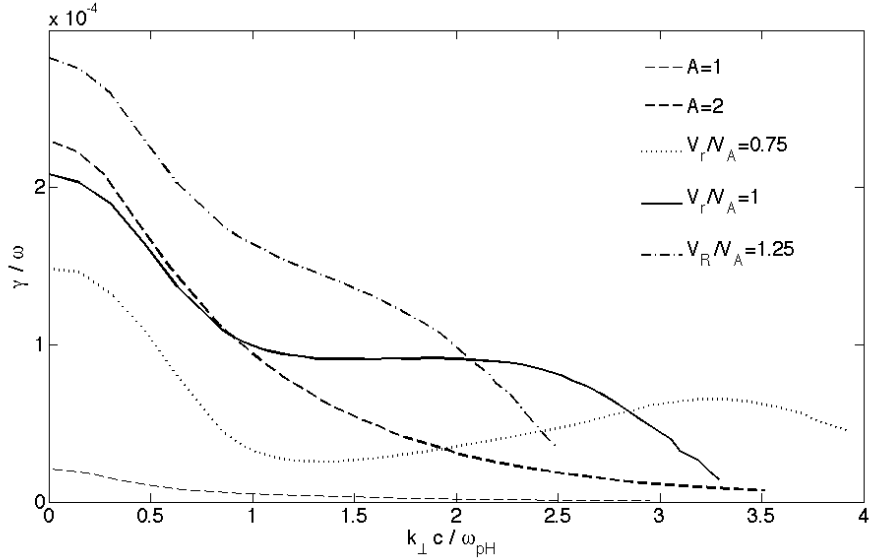

Figure 1. The maximum linear growth rate for EMIC waves due to a velocity-ring and a bi-Maxwellian distribution of hydrogen as a function of perpendicular wavenumber with $V_A^2 / v_{tr\|}^2 = 2$. The growth rate has been normalized to the unstable frequency and the wavenumber normalized to the ion skin depth. The ring density to electron density in this figure is $n_r / n_e = 10^{-3}$. Since the growth rate scales linear with $n_r / n_e$, other growth rates are easily obtained. Several values of $V_R / V_A$ are shown and the trend indicates that smaller ring speeds have smaller growth rates, but have larger growth rates at larger $k_\perp$. The larger growth rates at larger $k_\perp$ occur because the maximum of $-\partial J_1^2(\sigma) / \partial \sigma^2$ occurs near $(k_\perp c / \omega_{pH}) V_r / V_A \sim 2.3$.

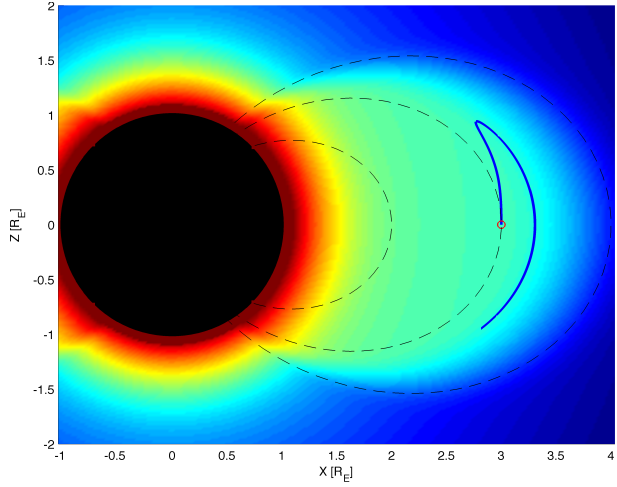

Figure 2. The density profile used for raytracing [from Bortnik et al., 2011]. The ray path, initialized at L=3, is traced over the contours of constant density used in the model calculation.

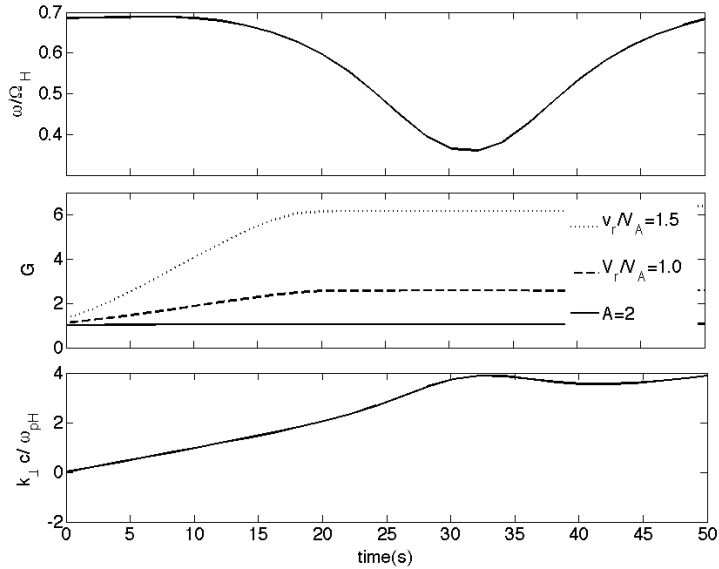

Figure 3. *Path Integrated wave gain*. (a) The frequency of the EMIC wave normalized to the hydrogen-cyclotron frequency as a function of time. The normalized frequency $\omega/\Omega_H$ decreases because the magnetic field decreases as the wave packet moves away from the equator. (b) The wave gain as a function of time. $V_A^2/v_{tr\parallel}^2 = 10$. The gain achieved from a velocity-ring distribution is several times larger than that of an anistropic bi-Maxwellian. The gain from the bi-Maxwellian is virtually negligible. Most of the gain is achieved during its initial transit away from the equator, though the gain from a ring distribution occurs over a much longer time-period since the growth rate is larger at oblique angles. (c) Perpendicular wave number. As the ray refracts, $k_\perp$ increases, and the wave-normal angle tends toward $90^o$.